\documentclass[12pt,thmsa,a4paper]{article}
\usepackage{sw20lart}

\input{tcilatex}
\input tcilatex
\QQQ{Language}{
American English
}

\begin{document}

\author{Luis Masperi \thanks{%
Present address: Centro Latinoamericano de F\'isica, Av. Wenceslau Br\'az 71
Fundos, 22290 - 140 R\'io de Janeiro, Brazil.} \thanks{%
E - mail : masperi@cbpfsu1.cat.cbpf.br} \\
%EndAName
Centro At\'omico Bariloche and Instituto Balseiro,\\
Comisi\'on Nacional de Energ\'\i a At\'omica and Universidad Nacional de
Cuyo,\\
8400 San Carlos de Bariloche, Argentina \and Milva Orsaria \thanks{%
E - mail : orsaria@tandar.cnea.edu.ar} \\
%EndAName
Laboratorio TANDAR,\\
Comisi\'on Nacional de Energ\'\i a At\'omica, Av. del Libertador 8250,\\
1429 Buenos Aires, Argentina}
\title{BARYOGENESIS THROUGH GRADUAL COLLAPSE OF VORTONS}
\maketitle

\begin{abstract}
We evaluate the matter-antimatter asymmetry produced by emission of
fermionic carriers from vortons which are assumed to be destabilized at the
electroweak phase transition. The velocity of contraction of the vorton,
calculated through the decrease of its magnetic energy, originates a
chemical potential which allows a baryogenesis of the order of the observed
value. This asymmetry is not diluted by reheating if the collapse of vortons
is distributed along an interval of $\sim 10^{-9}$sec.
\end{abstract}

\section{Introduction}

The matter-antimatter asymmetry in the universe is one of the well
established facts of cosmology. There are many possible mechanisms to
generate this baryonic density due to phenomena which presumably occurred in
the first fraction of second after the big-bang but all of them suffer some
criticism. They include also methods involving cosmic strings or other
topological defects produced in some of the phase transitions produced in
the universe.

In this work we present a baryogenesis model based on possibly very abundant
closed cosmic strings called vortons stabilized by superconducting currents,
which might lose this stability at the electroweak phase transition. The
distinctive feature of our mechanism is that we follow the process of
contraction of vortons and that we assume that they do not destabilize all
at the same time.

In Section 2 we give a survey of the baryogenesis methods which have some
connection with our proposal. In Section 3 we briefly describe vortons and
their relationship with Grand Unified Theories (GUT). Section 4 reminds the
instantaneous decay of vortons and the reheating which causes the dilution
of matter-antimatter asymmetry. In Section 5 we present our scenario of
gradual decay of vortons indicating how the reheating problem may be solved.
Section 6 contains the details of the calculation of the contraction
velocity which, for the case of charged carriers, is based on the decrease
of the associate magnetic energy. In Section 7 we evaluate by tunneling the
probability of emission of carriers which gives way to the asymmetry
produced by each vorton. Section 8 shows how the variation of magnetic field
due to contraction produces a chemical potential which allows our
baryogenesis to be of the order of the expected one. Section 9 contains some
conclusions.

\section{Methods to generate matter - antimatter asymmetry}

From the nucleosynthesis of light elements there is a costraint for the
baryonic density which, related to entropy density to give an invariant
value, is \cite{Kolb}

\begin{equation}
\frac{n_B}s=10^{-11}-10^{-10}\text{ . }  \label{e1}
\end{equation}

To explain this asymmetry, if one starts from a symmetric universe, three
conditions are required \cite{A.D.Sakharov} : i ) non-conservation of
baryonic number, ii ) violation of C and CP to distinguish particle from
antiparticle, iii ) period of non-equilibrium to allow different number of
particles and antiparticles.

The method of baryogenesis closest to experimental verification is that
which corresponds to the electroweak phase transition \cite{A.D.Dolgov}
provided it is a first-order one. Expanding bubbles of the broken-symmetry
phase would produce in its wall a chemical potential due to the variation of
a CP violating phase $\theta $ compared to the external symmetric medium,
where the active sphalerons would generate the baryonic density. The latter
would not be erased because the bubble expansion would include it in the
broken-symmetry phase where sphaleron processes are very slow. Due to the
rate of sphaleron transitions in the high-temperature phase one would obtain

\begin{equation}
\frac{n_B}s\simeq \frac{\alpha _w^4}{g^{*}}\text{ }\Delta \theta \text{ , }
\label{e2}
\end{equation}
where the weak coupling is such that $\alpha _w\simeq 10^{-3}$ and the
number of zero-mass modes at the electroweak temperature $T\sim 100GeV$ is $%
g^{*}\simeq 100.$ Therefore the observed asymmetry is reproduced if $\Delta
\theta \sim 10^{-2}.$ However this mechanism with Standard Model ingredients
is not possible because the phase transition turns out to be of second order
for the experimental bound on the Higgs mass and the CP violation is not
enough.

A solution which would include not too high-energy elements beyond the
Standard Model is afforded by the Minimal Supersymmetric Standard Model. But
this model would be severely constrained because to give an enough
first-order phase transition the Higgs boson should be light $m_H<100GeV$as
well as the stop $m_{\widetilde{t}}\leq 200GeV$, and to allow a large enough
variation of the CP violation parameter without entering in conflict with
the neutron electric dipole moment the lower generations of squarks should
be very heavy\cite{M.Carena}, though this last condition might be relaxed%
\cite{A.Riotto}.

On the other extreme of the energy range, a possibility of baryogenesis
would be given by the decay of GUT Higgs and gauge bosons which should be
produced out of equilibrium requiring $T\sim 10^{16}GeV$. The generated
baryonic density would be

\begin{equation}
\frac{n_B}s\simeq \frac \varepsilon {g^{*}}  \label{e3}
\end{equation}
where, with the asymmetry produced by one of these superheavy particles $%
\varepsilon \sim 10^{-8},$ there would be agreement with the expected value.

A problem is here that at these extremely high temperatures magnetic
monopoles would have been produced with the consecuent overclosure of the
universe density, as well as very heavy cosmic strings which might originate
undesirable inhomogeneities. It is anyhow difficult to explain such high T
from the reheating at the end of inflation, unless the non-linear quantum
effects of preheating give way to an explosive heavy particle production out
of equilibrium\cite{E.W.Kolb}.

\section{Cosmic strings and vortons}

Cosmic strings are topological defects which appear in a phase transition
when an abelian symmetry additional to the standard model is broken. To
avoid the monopole problem we may assume that the universe reached a
temperature for which the GUT symmetry G was already broken

\begin{equation}
G\rightarrow SU\left( 3\right) _C\times SU\left( 2\right) _L\times U\left(
1\right) _Y\times \widetilde{U}\left( 1\right) \text{ .}  \label{e4}
\end{equation}

If at a slightly lower temperature, let us say $10^{15}GeV$, also the
symmetry $\widetilde{U}\left( 1\right) $ is broken the cosmic strings will
be produced\cite{T.W.B.Kibble}. They will become superconducting\cite
{E.Witten} depending on the group G and the details of the Higgs mechanism
for the breaking of $\widetilde{U}\left( 1\right) $. A superconducting
current will appear for those fermionic carriers which acquire mass due to
the coupling with a Higgs field which winds the string and originates zero
modes inside it. The superconducting current classically stabilizes closed
loops through a number $N$ related to the angular momentum due to the
carriers inside them.

It is not necessary that the carriers are charged\cite{R.L.Davis}. In fact
if $G=SO\left( 10\right) $ the only particle which acquires mass at the $%
\widetilde{U}\left( 1\right) $ phase transition is $\nu _R$ which may have a
zero mode inside the string. On the other hand if $G=E_6$ several fermions
acquire mass at the $\widetilde{U}\left( 1\right) $ breaking and some of
them, which may give superconducting currents, are charged and with baryonic
number.

For normal cosmic strings it is interesting\cite{P.Bhattacharjee} that if
the emission of superheavy bosons at present time is normalized to explain
the flux of ultra-high energy cosmic rays, their decay in the past may give
the expected baryogenesis provided that the asymmetry per particle is six
orders of magnitude higher than that necessary in Eq.(3).

The stabilized superconducting closed loops are called vortons\cite{Shellard}%
. Their number density, mass, length and quantum decay probability depend on
the coincidence or not of the scales of string formation and appearance of
superconductivity in them\cite{R.Brandenberger}. If both scales coincide at $%
m_x$ the vorton density is

\begin{equation}
n_v\simeq \left( \frac{m_x}{m_{pl}}\right) ^{\frac 32}T\text{ }^{\text{3}}%
\text{ , }  \label{e5}
\end{equation}
its energy $E_v\simeq N$ $m_x$, radius $R\simeq $ $N$ $m_x^{-1}$ and $N\sim
10$ if $m_x\sim 10^{15}GeV$. If the superconductivity scale is smaller than
the formation one, the density is smaller and vortons are more stable for
quantum decay.

The density Eq.(5) overcloses the universe in a way similar to that of
monopoles, if there is not a collapse of vortons for some reason. If this is
produced at high energy when the carriers are $\nu _R$, a lepton asymmetry
appears which may be converted into baryon asymmetry by sphaleron processes
to give the expected value with adequately large CP violation parameter\cite
{R.Jeannerot}. Alternatively, if superconductivity appears at much lower
temperature, i.e. at the supersymmetric scale $\sim 1TeV$, there are models
predicting that vortons which subsequently decay below the electroweak
temperature may release baryonic charge in agreement with the expected one%
\cite{Riotto}, again assuming an adequate CP violation factor.

It must be noted that if the scales of formation and superconductivity
coincide and the vorton density is decreased for some process to be
constrained by the critical density of universe, the quantum decay
probability might be enough to explain the high energy cosmic rays\cite
{L.Masperi}.

\section{Instantaneous decay of vortons at the electroweak transition}

Trying to include as few ingredients as possible, we will adopt the point of
view that vortons have obtained the superconducting property at the same
scale of formation, and that they lose their stability at the well
established electroweak transition. This may occur if, due to the new Higgs
mechanism at this scale, the zero modes acquire a small mass\cite{S.C.Davis}%
. It is not required that the transition is of first order.

If vortons disappear instantaneously, since they contain roughly N heavy
bosons the produced baryonic density is

\begin{equation}
\frac{n_B}s=\left( \frac{m_x}{m_{pl}}\right) ^{\frac 32}\frac{N\varepsilon }{%
g^{*}}\text{ , }  \label{e6}
\end{equation}
which will be very small if the asymmetry due to each particle X is of the
same order of that of GUT bosons assumed in Eq.(3).

Furthermore, since vortons behave as non-relativistic matter, its density
which is very small at formation becomes equivalent to that of radiation at $%
T\sim 10^8GeV$ and dominates on it by 6 orders of magnitude at the
electroweak scale. Therefore if at this temperature vortons transform
instantaneously into light particles, i.e.radiation, there will be a
reheating according to

\begin{equation}
\rho _v\left( T_{EW}\right) =N\text{ }m_x\left( \frac{m_x}{m_{pl}}\right)
^{\frac 32}T_{EW}^{\text{3}}=\rho _{_R}\left( T_{reh}\right) =g^{*}T_{reh}^4%
\text{ ,}  \label{e7}
\end{equation}
which gives $T_{reh}\simeq 10^{\frac 72}GeV$.

This instantaneous reheating would produce an increase of the entropy
density of $\left. \left( \frac{Treh}{T_{EW}}\right) ^3\simeq 10^{\frac
92}\right. $ times with the corresponding dilution \cite{W.B.Perkins} of the
baryogenesis of Eq.(6) in the same factor.

According to this scenario the universe would be initially dominated by
radiation, then from $T\sim 10^8GeV$ to $T_{EW}$ by vortons and after the
reheating to $T_{reh}$ again by radiation till $t\sim 10^{11}\sec .$ when
finally non-relativistic matter takes over.

\section{Gradual collapse of vortons}

The alternative that we wish to present corresponds to the plausible
situation that vortons are not destabilized all at the same time when
reaching the electroweak temperature. Due to the Higgs mechanism that will
be working in this phase transition, we expect a probability that a vorton
loses its zero modes and starts its collapse. Without attempting to
calculate this probability for destabilization, we remark that the
temperature will remain constant at $T_{EW}\sim 100GeV$ if vortons decay
during an interval such that the universe expands its scale from $a_1$ to $%
a_2$ when all is transformed to radiation

\begin{equation}
a_{1}^{3}\text{ }N\text{ }m_{x}\left( \frac{m_{x}}{m_{pl}}\right) ^{\frac{3}{%
2}}T_{EW}^{3}=a_{2}^{3}\text{ }g^{*}T_{EW}^{4}\text{ .}  \label{e8}
\end{equation}

The space scale would therefore increase in two orders of magnitude and,
using $\frac{a_1}{a_2}=\left( \frac{t_1}{t_2}\right) ^{\frac 23}$, if the
process starts at $t_1\sim 10^{-12}\sec $ it would be completed at $t_2\sim
10^{-9}\sec $. The advantage is now that the total increase of entropy,
which is similar to that of instantaneous destabilization, is distributed in
a larger volume. Baryogenesis would not be diluted at the beginning of the
interval but only at the end with a factor $\left( \frac{a_2}{a_1}\right) ^3$%
so that the average dilution would be $\sim \frac 12.$ It is reasonable to
think that the collapse of vortons keeps the temperature constant because as
soon as there is a tendence to reheating the symmetry is restored and the
destabilization of vortons stops.

Furthermore, we will follow the contraction of each vorton obtaining
baryogenesis not by the presumably small asymmetry in the decay of bosons X,
but from the emission of charged baryonic carriers during the collapse. The
resulting asymmetry per vorton may turn out to be larger due to the chemical
potential which will appear in the wall of the vorton because of the
non-equilibrium process of contraction, resulting in a different emission of
fermions and antifermions.

\section{Velocity of contraction during vorton decay}

The evaluation of the velocity of contraction of the vorton after its
destabilization at the electroweak temperature is crucial for determining
the non-equilibrium process.

One possibility of calculation, which may be applied to the case of neutral
carriers, is to consider that stabilization is abruptly lost at $T_{EW}$ so
that the string contracts due to a constant tension $\mu \sim m_x^2$. If the
string mass were constant and the initial radius is $R\sim \frac N{m_x}$,
the relation between the velocity and each radius r would be

\begin{equation}
\text{v}^2\simeq 2\text{ }\frac{R-r}R\text{ .}  \label{e9}
\end{equation}

Considering that the vorton mass decreases linearly with its radius in the
rest frame, including the Lorentz factor and being at the initial stage of
the contraction when the iterative approximation may be used, the velocity
turns out to be 
\begin{eqnarray}
\text{v} &=&\frac 1{\sqrt{2}}\arctan \frac{\sqrt{2}N}{m_xt}\left[ 1-\left( 
\frac{m_xt}N\right) ^2\right] ^{\frac 12}+2\arcsin \left( \frac{m_xt}N\right)
\label{e10} \\
&&-\frac 3{2\sqrt{2}}\ln \left( \left| \frac{m_xt+\sqrt{2}N\left[ 1-\left( 
\frac{m_xt}N\right) ^2\right] ^{\frac 12}}{m_xt-\sqrt{2}N\left[ 1-\left( 
\frac{m_xt}N\right) ^2\right] ^{\frac 12}}\right| \right) -\frac \pi {2\sqrt{%
2}}\text{ .}  \nonumber
\end{eqnarray}

To have the relation between velocity and radius which replaces Eq.(9), v of
Eq.(10) should be integrated on time. It is clear that v will vary from 0 to
1 when r = 0 so that the time of collapse will be $t_{c}\geq \frac{N}{m_{x}}$
.

For charged carriers, in which we are more interested, the velocity of
contraction may be calculated in an easier way. We consider the decay of a
vorton as a succession of transitions between superconducting states of
numbers N, N-1,...keeping the value of the current I. Looking at a classical
average, one will see a loop with increasing contraction velocity with the
corresponding relativistic factor in its mass which will be compensated by
the decrease of the magnetic energy that is defined in the broken-symmetry
phase.

The balance for the vorton when its radius is r and the associated magnetic
field is B compared with the initial one B$_i,$ will be

\begin{equation}
\frac{r}{R}\text{ }N\text{ }m_{x}\left( \frac{1}{\sqrt{1-\text{v}^{\text{2}}}%
}-1\right) =\frac{1}{2}\int d\mathbf{\rho }\left( B_{i}^{2}-B^{2}\right) 
\text{ }.  \label{e11}
\end{equation}

Since in general we will expect

\begin{equation}
\frac{1}{2}\int d\mathbf{\rho }\text{ }B^{2}=k\text{ }I^{2}\text{ }r
\label{e12}
\end{equation}

and being $I=\frac{N}{2\pi R}$ , we will have

\begin{equation}
1-\text{v}^2=\frac 1{\left( 1+\frac{k_i\text{ }R-k_f\text{ }r}{4\pi ^2\text{ 
}r}\right) ^2}\text{ ,}  \label{e13}
\end{equation}

where the coefficient k$_{f}$ at the end of the collapse may be different
from the initial one k$_{i}$.

At the beginning, when $R-r$ is small and $k_i=k_f$

\begin{equation}
\text{v}^2\simeq \frac{k_i}{2\pi ^2}\text{ }\frac{R-r}R\text{ ,}  \label{e14}
\end{equation}
which is analogous to the previous estimation due to constant tension. For $%
r\rightarrow 0$ the velocity Eq.(13) will tend to 1.

An important ingredient for our evaluation of baryogenesis will be the
calculation of the coefficients k.

In the first part of the contraction the vorton will be certainly well
represented by a loop of radius r that, lying in the x-y plane, will give a
magnetic potential

\begin{equation}
A_\varphi \left( \rho ,\theta \right) =I\text{ }r\text{ }\int_0^{2\pi
}d\varphi ^{\prime }\frac{\cos \varphi ^{\prime }}{\left( \rho ^2+r^2-2r\rho
\sin \theta \cos \varphi ^{\prime }\right) ^{\frac 12}}\text{ .}  \label{e15}
\end{equation}

For large distances $\rho \gg r$ Eq.(15) gives the dipole approximation for
the magnetic field

\begin{equation}
B_\rho =\frac{2m\cos \theta }{\rho ^3}\text{ , }B_\theta =\frac{m\sin \theta 
}{\rho ^3}\text{ ,}  \label{e16}
\end{equation}
with $m=\pi r^2I$ .

For distances much smaller than the radius $\rho \ll r$

\begin{equation}
B_\rho =\frac{2\pi I}r\text{ }\cos \theta \text{ , }B_\theta =-\text{ }\frac{%
2\pi I}r\text{ }\sin \theta \text{ ,}  \label{e17}
\end{equation}
whereas the exact expressions from Eq.(15) are

\begin{eqnarray*}
B_{\rho } &=&\cot \theta \text{ }I\text{ }\frac{r}{\rho }\text{ }%
\int_{0}^{2\pi }d\varphi ^{\prime }\text{ }\frac{\cos \varphi ^{\prime }}{%
\left( \rho ^{2}+r^{2}-2r\rho \sin \theta \cos \varphi ^{\prime }\right) ^{%
\frac{1}{2}}} \\
&&+I\text{ }r^{2}\cos \theta \text{ }\int_{0}^{2\pi }d\varphi ^{\prime }%
\text{ }\frac{\cos ^{2}\varphi ^{\prime }}{\left( \rho ^{2}+r^{2}-2r\rho
\sin \theta \cos \varphi ^{\prime }\right) ^{\frac{3}{2}}}
\end{eqnarray*}

\begin{eqnarray}
B_{\theta } &=&-I\text{ }\frac{r}{\rho }\text{ }\int_{0}^{2\pi }d\varphi
^{\prime }\text{ }\frac{\cos \varphi ^{\prime }}{\left( \rho
^{2}+r^{2}-2r\rho \sin \theta \cos \varphi ^{\prime }\right) ^{\frac{1}{2}}}
\label{e18} \\
&&+I\text{ }r\text{ }\rho \text{ }\int_{0}^{2\pi }d\varphi ^{\prime }\text{ }%
\frac{\cos \varphi ^{\prime }}{\left( \rho ^{2}+r^{2}-2r\rho \sin \theta
\cos \varphi ^{\prime }\right) ^{\frac{3}{2}}}  \nonumber \\
&&-I\text{ }r^{2}\sin \theta \text{ }\int_{0}^{2\pi }d\varphi ^{\prime }%
\text{ }\frac{\cos ^{2}\varphi ^{\prime }}{\left( \rho ^{2}+r^{2}-2r\rho
\sin \theta \cos \varphi ^{\prime }\right) ^{\frac{3}{2}}}\text{ .} 
\nonumber
\end{eqnarray}

For small $\sin \theta $, Eq.(18) may be approximated by

\begin{equation}
B_\rho =\frac{2\pi r^2I}{\left( \rho ^2+r^2\right) ^{\frac 32}}\text{ }\cos
\theta \text{ , }B_\theta =\frac{\pi r^2I}{\left( \rho ^2+r^2\right) ^{\frac
32}}\text{ }\frac{\rho ^2-2r^2}{\rho ^2+r^2}\text{ }\sin \theta \text{ .}
\label{e19}
\end{equation}

We may evaluate the magnetic energy, except for the x-y plane, taking the
dipole approximation Eq.(16) for $\rho >3$ $r$, the small $\rho $
approximation Eq.(17) for $\rho <\frac r3$ , and the intermediate expression
Eq.(19) for $\frac r3<\rho <3$ $r$ since it matches well with the other ones
at these values. In this way one obtains a contribution to k in Eq.(12) of $%
3.5\pi ^3$ , but it is still necessary to add the contribution near the
plane x-y.

For this last part, the contribution will come essentially from the region
close to the loop. For $\theta =\frac \pi 2$ , $B_\theta $ of Eq.(18) will
have a logaritmic divergence for $\rho =r$ coming from $\varphi ^{\prime
}\sim 0$, which is regularized considering the width of the loop and that
inside it $B=0$ to be a superconducting medium.

With the approximation of keeping a region $\alpha $ of integration of $%
\varphi ^{\prime }$ near to zero and comparing the finite contribution to $%
B_\theta $ with that coming from the exact evaluation of Eq. (18) for $\rho
=r$, its turns out that $\alpha \simeq 0.77$ . Taking now for $\rho =r+\eta $
the approximation of keeping up to terms $\varphi ^{\prime \text{ }2}$ in
Eq.(18), for $\frac{\left| \eta \right| }r<<1$, one obtains

\begin{equation}
B_\theta \simeq \frac 1{\left( \alpha ^2r^2+\eta ^2\right) ^{\frac 12}}\text{
}\frac{2\alpha Ir}\eta +\left[ \ln \frac{\left| \eta \right| }r-\ln \left(
\alpha +\sqrt{\alpha ^2+\frac{\eta ^2}{r^2}}\right) \right] \frac Ir\text{ . 
}  \label{e20}
\end{equation}

A reasonable estimation of the magnetic energy near the string, considering
that it must correspond to the stages of contraction starting from $R\sim
\frac N{m_X\text{ }}$ and being $\eta \sim \frac 1{m_X}$ , is to take the
above approximation for $B_\theta $ in an external region of size $\eta $
around it. It turns out that this contribution to the coefficient $k$ will
be $\sim 2\pi ^3.$

Therefore the total contribution of magnetic energy when the decaying vorton
may still be considered as a thick loop corresponds to

\begin{equation}
k_i\simeq 5.5\pi ^3\text{ . }  \label{e21}
\end{equation}

With this value, the expression Eq.(14) for the velocity of contraction
using the magnetic energy is of the same order of that given by Eq.(9) for
the initial stage with constant string tension.

But it is not always correct to consider the contracting vorton as a loop.
At the final stage when the radius is of the order of the width it may be
better to represent it as a sphere with currents running inside it around
the z-axis. Approximating each disc at angle $\theta ^{\prime }$ by an
effective loop of radius $\eta \sin \theta ^{\prime }$, the magnetic
potential will be

\begin{equation}
A_\varphi \left( \rho ,\theta \right) =I\eta \int_0^\pi d\theta ^{\prime
}\sin \theta ^{\prime }\int_0^{2\pi }d\varphi ^{\prime }\frac{\cos \varphi
^{\prime }}{\left[ \rho ^2+\eta ^2-2\rho \eta \left( \sin \theta \sin \theta
^{\prime }\cos \varphi ^{\prime }+\cos \theta \cos \theta ^{\prime }\right)
\right] ^{\frac 12}}\text{ .}  \label{e22}
\end{equation}

For large distances $\rho >>\eta $ , Eq.(22) gives again the dipole limit

\begin{equation}
A_\varphi \left( \rho ,\theta \right) =I\frac{\pi \eta ^2}{\rho ^2}\text{ }%
\frac \pi 2\text{ }\sin \theta \text{ . }  \label{e23}
\end{equation}

On the other hand, near the sphere $\rho =\eta +\delta $ the contribution to
the magnetic field will come mainly from a region $\alpha $ in the
integration over $\varphi ^{\prime }$ for small values such that only terms $%
\varphi ^{\prime }$ $^2$are kept and a region $\beta $ in the integration
over $\theta ^{\prime }$ such that $\theta ^{\prime }\simeq \theta $. The
result is

\begin{eqnarray*}
B_\rho \left( \delta ,\theta \right) &=&2\beta \cos \theta \text{ }I\eta 
\text{ }\left[ \frac 1{2c\left( \eta +\delta \right) }\text{ }\left( 1+\frac{%
\delta ^2}{2c^2}\right) \ln \left( \frac{\alpha c}\delta +\sqrt{1+\frac{%
\alpha ^2c^2}{\delta ^2}}\right) \right. \\
&&\left. -\frac 1{\eta +\delta }\text{ }\frac{\alpha \delta }{4c^2}\sqrt{1+%
\frac{\alpha ^2c^2}{\delta ^2}}+\frac 12\text{ }\frac{\alpha \eta }{\delta
c^2}\text{ }\frac{\sin ^2\theta }{\sqrt{1+\frac{\alpha ^2c^2}{\delta ^2}}}%
\right] \text{ ,}
\end{eqnarray*}

\begin{eqnarray}
B_\theta \left( \delta ,\theta \right) &=&2\beta \sin \theta \text{ }I\eta
\left\{ -\left[ \frac 1{2c\left( \eta +\delta \right) }\left( 1+\frac{\delta
^2}{2c^2}\right) +\frac \delta {2c^3}\right] \ln \left( \frac{\alpha c}%
\delta +\sqrt{1+\frac{\alpha ^2c^2}{\delta ^2}}\right) \right.  \label{e24}
\\
&&\left. +\frac 1{\eta +\delta }\text{ }\frac{\alpha \delta }{4c^2}\text{ }%
\sqrt{1+\frac{\alpha ^2c^2}{\delta ^2}}-\frac 12\text{ }\frac{\alpha \eta }{%
\delta c^2}\text{ }\frac{\sin ^2\theta }{\sqrt{1+\frac{\alpha ^2c^2}{\delta
^2}}}\text{ }+\frac \alpha {\sqrt{1+\frac{\alpha ^2c^2}{\delta ^2}}}\left(
\frac 1{\delta ^2}+\frac 1{2c^2}\right) \right\} \text{ ,}  \nonumber
\end{eqnarray}
where $c=\sqrt{\eta ^2+\eta \delta }\sin \theta $ .

The limit of Eq.(24) for small values of $\sin \theta $ is

\begin{equation}
B_\rho \rightarrow 2\beta \alpha \cos \theta \frac{I\eta }{\delta \rho }%
\text{ },\text{ }B_\theta \rightarrow 2\beta \alpha \sin \theta \frac{I\eta
^2}{\rho \delta ^2}\text{ , }  \label{e25}
\end{equation}
whereas for $\theta \sim \frac \pi 2$ and $\alpha \sim 0.77$ , $B_\rho \sim
0 $ and a numerical evaluation of Eq.(24) gives

\begin{equation}
B_\theta \left( \delta ,\theta =\frac \pi 2\right) \simeq 0.35\text{ }\beta 
\text{ }\frac I\delta \text{ .}  \label{e26}
\end{equation}

We now take two regions for evaluating the magnetic energy : that for large $%
\rho $ where the field corresponds to the dipole approximation and that
close to the sphere, since inside it the field is zero for a superconducting
medium. The two regions match reasonably well for $\delta =\eta $ and the
lower limit for the integration is $\rho =\frac 32$ $\eta $ because $\eta $
was the average radius of the disc in the x-y plane.

Therefore the coefficient for the magnetic energy when the vorton is
approximated by a sphere turns out to be, with $\beta \sim \alpha $,

\begin{equation}
k_f\simeq 35\text{ .}  \label{e27}
\end{equation}

\section{Probability of emission of carriers}

We will calculate the matter-antimatter asymmetry per vorton through the
emission of fermions and antifermions by quantum tunneling. This corresponds
to the transition e.g. from a state of vorton with number $N$ to another
with number $N-1$ plus a fermion of mass $m_x$ with conservation of angular
momentum.

It must be stressed that this channel is not the dominant one for the
contraction of the string since the corresponding partial lifetime is much
longer than the actual time of collapse. But it turns out to be the most
effective one for baryogenesis since, due to the chemical potential produced
by the non-equilibrium process of contraction, the probability for emission
of baryons will be substancially different from that of antibaryons. In
comparison, other channels which eliminate pieces of string due to the
destabilization produced at the electroweak transition will give through the
decay of heavy bosons a rather small amount of matter-antimatter asymmetry
as discussed in Section 4.

We evaluate the tunneling process semiclassically. The height of the barrier
will be of the order of $m_x$ since it corresponds to the increase of energy
when the massless carrier inside the string is put outside it with the same
momentum. Additionally, the width of the barrier is the displacement of the
carrier such that, always conserving angular momentum and taking into
account the one - step contraction of the string, the energy of the
configuration is equal to the initial one. This displacement turns out to be
of the order of the radius r of the emitting string \cite{L.Masperi}.

Therefore, the emission probability in the string rest frame will be

\begin{equation}
\Gamma _0\simeq m_x^2\text{ }re^{-m_xr}\text{ .}  \label{e28}
\end{equation}

Considering that the probability in laboratory frame requires the
relativistic factor for the dilatation of time

\begin{equation}
\Gamma =\Gamma _0\sqrt{1-\text{v}^2}\text{ ,}  \label{e29}
\end{equation}
and that the difference between emission of particle and antiparticle is
given by its multiplication times

\begin{equation}
-\frac \mu T=\text{v }\Delta \text{ ,}  \label{e30}
\end{equation}
where $\mu $ is the chemical potential and $\Delta $ will depend on a
specific contribution, the asymmetry due to a vorton during all the time of
its collapse will be

\begin{equation}
\varepsilon _v=\Delta \text{ }m_x^2\int_\eta ^Rdr\text{ }r\text{ }\sqrt{1-%
\text{v}^2}\text{ }e^{-m_xr}\text{ . }  \label{e31}
\end{equation}

Defining $y=m_x$ $r$ , and being $R\simeq \frac N{m_x}$ and $\eta \simeq
\frac 1{m_x}$, from Eq.(13) one has

\begin{equation}
\varepsilon _v=\Delta \int_1^Ndy\frac{y^2\text{ }e^{-y}}{\frac{k_i}{4\pi ^2}%
N+\left( 1-\frac{k_f}{4\pi ^2}\right) y}\text{ . }  \label{e32}
\end{equation}

Due to the fact that $k_i$ corresponds always to the loop approximation of
vorton but $k_f$ may correspond either to loop or to sphere approximations,
the integral of Eq.(32) must be splitted into two parts

\begin{equation}
\frac{\varepsilon _v}\Delta =\int_{N_s}^Ndy\text{ }\frac{y^2\text{ }e^{-y}}{%
\widetilde{k}N+\left( 1-\widetilde{k}\right) y}+\int_1^{N_s}dy\text{ }\frac{%
y^2\text{ }e^{-y}}{\widetilde{k}N+\left( 1-\widetilde{k}_f\right) y}=I_1+I_2%
\text{ , }  \label{e33}
\end{equation}
where we have called $\widetilde{k}\simeq 5$ and $\widetilde{k}_f\simeq 1$
according to Eqs. (21) and (27).

These integrals can be done exactly but it is instructive also to calculate
them expanding the denominators in powers of $\frac{\left( \widetilde{k}%
-1\right) y}{\widetilde{k}N}<1$ and $\frac{\left( \widetilde{k}_f-1\right) y%
}{\widetilde{k}N}<1$ giving

\begin{equation}
I_1=\frac 1{\widetilde{k}N}\text{ }\sum_{n=0}^\infty \left( \frac{\widetilde{%
k}-1}{\widetilde{k}N}\right) ^n\left. \left\{ \left( n+2\right) !-\left[
y^{n+2}+\left( n+2\right) \text{ }y^{n+1}+...\left( n+2\right) !\right]
e^{-y}\right\} \right| _{N_s}^N  \label{e34}
\end{equation}
and for $I_2$ a similar expression where the coefficient in the numerator is 
$\widetilde{k}_f-1$ and the limits 1 and $N_s$.

For order $n=0$ there is no influence of the difference between $\widetilde{k%
}$ and $\widetilde{k}_f$ and of the value of $N_s$.

\begin{equation}
\frac{\varepsilon _v^{(0)}}\Delta =\frac 1{\widetilde{k}N}\left[ \frac
5e-\left( N^2+2N+2\right) \text{ }e^{-N}\right] \simeq 0.03\text{ . }
\label{e35}
\end{equation}
The contribution of $n=1$ adds, taking $N_s=2$,

\begin{eqnarray}
\frac{\varepsilon _v^{(1)}}\Delta &=&\frac 1{\left( \widetilde{k}N\right)
^2}\left[ \left( \widetilde{k}_f-1\right) \frac{16}e+\left( \widetilde{k}-%
\widetilde{k}_f\right) \left( N_s^3+3N_s^2+6N_s+6\right) e^{-N_s}\right.
\label{e36} \\
&&\left. -\left( \widetilde{k}-1\right) \left( N^3+3N^2+6N+6\right)
e^{-N}\right]  \nonumber \\
&\simeq &0.005\text{ .}  \nonumber
\end{eqnarray}
Therefore we may expect $\frac{\varepsilon _v^{}}\Delta $ close to $0.1$ .

In fact the exact evaluation of the asymmetry per vorton gives

\begin{eqnarray}
I_1 &=&\left[ \frac{\left( \gamma y+\widetilde{k}N\right) ^2}{2\gamma ^3}-%
\frac{2\widetilde{k}N}{\gamma ^3}\left( \gamma y+\widetilde{k}N\right) +%
\frac{\left( \widetilde{k}N\right) ^2}{\gamma ^3}\ln \left( \gamma y+%
\widetilde{k}N\right) \right] e^{-y}  \label{e37} \\
&&-\frac{e^{-y}}{2\gamma ^3}\left\{ \left[ \gamma \left( y+1\right) +%
\widetilde{k}N\right] ^2-4\left( \widetilde{k}N\right) ^2+\gamma ^2\right\}
-e^{-y}\frac{\left( \widetilde{k}N\right) ^2}{\gamma ^3}\ln \left( \gamma y+%
\widetilde{k}N\right)  \nonumber \\
&&+\left. \frac{\left( \widetilde{k}N\right) ^2}{\gamma ^3}e^{\frac{%
\widetilde{k}N}\gamma }\ln \left( \gamma y+\widetilde{k}N\right) +\frac{%
\left( \widetilde{k}N\right) ^2}{\gamma ^3}e^{\frac{\widetilde{k}N}\gamma
}\sum_{n=1}^\infty \frac{\left( -1\right) ^n}{nn!}\left( y+\frac{\widetilde{k%
}N}\gamma \right) ^n\right| _{N_s}^N\text{ , }  \nonumber
\end{eqnarray}
where $\gamma =1-\widetilde{k}$, and a similar expression for $I_2$ with $%
\gamma =1-\widetilde{k_f}$ and the limits 1 and $N_s$.

The numerical computation with the above values of $\widetilde{k}$ and$%
\widetilde{\text{ }k_f}$ and $N_s=2$ gives $I_1=0.0487$ and $I_2=0.0078$ so
that

\begin{equation}
\frac{\varepsilon _v}\Delta =0.0565\text{ .}  \label{e38}
\end{equation}

All what we still need to calculate is the chemical potential to have the
numerical value of $\Delta $.

It must be added that we assume that inside the string one has the
high-temperature phase in thermal equilibrium so that there
matter-antimatter symmetry is kept.

\section{Chemical potential}

In the outer part of the string, the non-equilibrium process of contraction
will produce a chemical potential which should be otherwise zero for a
non-conserved charge as the baryonic one.

In our case the chemical potential may have two sources. One of them is
traditional as it appears in the expanding bubbles of electroweak
baryogenesis. In the Hamiltonian the baryonic density appears multiplied by $%
-\frac{d\theta }{dt}$ where $\theta $ is a CP violating phase which is
nonzero outside the string. During the contraction of the latter, points
which are crossed by its external wall pass $\theta $ from $0$ to a finite
value $\Delta \theta >0$ so that

\begin{equation}
\mu =-\frac{d\theta }{dt}=\frac{\Delta \theta }\eta \text{ v}<0\text{ ,}
\label{e39}
\end{equation}
and therefore emission of matter is favoured on antimatter. Since the
emission probability must be multiplied by $-\frac \mu {T\text{ }}$ and $%
\eta \sim \frac 1T$ in the high-temperature phase, in our expression of $%
\varepsilon _v$ Eq.(31) $\Delta =\Delta \theta $ which, as said in Section 2
should be $\sim 0.01$ to be in agreement with the bound of the electric
dipole moment of neutron. This contribution might be too small to give the
expected baryogenesis with our mechanism.

But our collapsing superconducting loop has another source of chemical
potential due to the magnetic field that it generates. Outside the external
wall of the string, which is where the emission occurs, these will be a
potential multiplying the fermionic density with charge q in the Hamiltonian

\begin{equation}
\mu =q\int_0^\varphi \frac \partial {\partial t}A_\varphi \text{ }r\text{ }%
d\varphi ^{\prime }\text{ , }  \label{e40}
\end{equation}
corresponding to the electric field generated by time variation of $%
A_\varphi $ due to the contraction of the loop.

It is interesting to note that this contribution to chemical potential can
be also thought as the difference of a phase if one thinks that in the wall
of the string the magnetic potential $A_\varphi $ will produce a change of
phase of the fermionic field which can be compensated by the transformation

\begin{equation}
\Psi \left( \varphi \right) \longrightarrow e^{iq\int_0^\varphi A_\varphi 
\text{ }dl\text{ }}\text{ }\Psi \left( \varphi \right) \text{ . }
\label{e41}
\end{equation}
But in so doing the kinetic term of the Dirac energy will acquire a
contribution of the tipe of $\mu $ Eq.(40) times the fermionic density due
to the time variation of $A_\varphi $ .

To evaluate the contribution of Eq.(40) one must calculate $A_\varphi $ of
Eq.(15) for $\rho =r+\eta $ and $\theta =\frac \pi 2$ and derivate it at
fixed $\rho $ with respect to time due to the variation of r. The most
important contribution to $A_\varphi $ is

\begin{equation}
A_\varphi \left( \rho =r+\eta ,\theta =\frac \pi 2\right) \simeq -2\text{ }I%
\text{ }\ln \left( \frac{\rho -r}r\right) \text{ ,}  \label{e42}
\end{equation}
so that

\begin{equation}
\left. \frac{\partial A_\varphi }{\partial t}\right| _\rho \simeq -2\frac
I\eta \text{ v , v}=-\frac{dr}{dt}\text{ .}  \label{e43}
\end{equation}
Because of the definition of Eq.(40), it will correspond to take an average
of the potential between $0$ and $2\pi $, i.e.

\begin{equation}
\left\langle \text{ }\mu \right\rangle =-q2\pi \frac{Ir}\eta \text{ v .}
\label{e44}
\end{equation}
Considering again the factor $-\frac \mu T$ which multiplies the emission
probability, we have

\begin{equation}
\Delta =q2\pi Ir\text{ .}  \label{e45}
\end{equation}
This coefficient will vary during the contraction. At the beginning $r\sim
R\simeq \frac N{m_x}$ and being the electric charge of the carrier $q\sim
0.1 $ and $I=\frac{m_x}{2\pi }$ it turns out that $\Delta \simeq 0.1$ ,
which is larger than the previous contribution to $\mu $ .

Therefore we obtain that $\varepsilon _v\sim 0.01$ and our asymmetry due to
gradual collapse of vortons will be

\begin{equation}
\frac{n_B}s\simeq \left( \frac{m_x}{m_{pl}}\right) ^{\frac 32}\frac{%
\varepsilon _v}{g*}\sim 10^{-10}\text{ .}  \label{e46}
\end{equation}
Considering that $\Delta $ may decrease towards the end of the collapse in
one order of magnitude and accepting a moderate dilution effect due to
gradual transformation of vortons into radiation as discussed in Section 5, $%
\frac{n_B}s$ might not go below $10^{-11}$ which is the lower limit of the
acceptable baryogenesis.

\section{Conclusions}

We have found that the emission of fermions from vortons destabilized at the
electroweak transition during their collapse may supply the
matter-antimatter asymmetry required by nucleosynthesis. This avoids on the
one hand the necessity of very high temperature of reheating after inflation
to produce superheavy bosons of GUT, and on the other the requirement of
first order for the electroweak transition needed for the production of
bubbles.

It is clear that our evaluation gives only a possible order of magnitude for
the baryogenesis with this mechanism. A more precise result would require a
calculation of the emission probability beyond the semiclassical approach,
as well as a detailed analysis of the disappearance of zero modes in vortons
to estimate the time interval for their destabilization.

It is interesting to note that if a particular Grand Unification model loses
not all its zero-modes at the electroweak temperature and a part of present
dark matter is due to vortons, their emission might explain the observed
high-energy cosmic rays, so that this phenomenon would be linked to that of
baryogenesis.

\begin{center}
\textbf{Acknowledgments}

This research was partially supported by CONICET PICT 0358. L.M. wishes to
thank the hospitality of the Physics Dept. of Universit\`a di Napoli and
Universit\`a della Calabria at Cosenza, where parts of this work were
performed.
\end{center}

\end{document}